\documentclass{aip-cp}

\usepackage[numbers]{natbib}
\usepackage{rotating}
\usepackage{graphicx}
\usepackage{siunitx}

\begin{document}

\title{Discovery of new TeV supernova remnant shells in the Galactic plane with H.E.S.S.}

\author[aff1]{D. Gottschall\corref{cor1}}
\author[aff1]{M. Capasso}
\author[aff2]{C. Deil}
\author[aff3]{A. Djannati-Atai}
\author[aff2]{A. Donath}
\author[aff2]{P. Eger}
\author[aff2]{V. Marandon}
\author[aff4]{N. Maxted}
\author[aff1]{G. P\"uhlhofer}
\author[aff5]{M. Renaud}
\author[aff1]{M. Sasaki}
\author[aff3]{R. Terrier}
\author[aff6]{J. Vink}
\author[]{for the H.E.S.S. Collaboration}

\affil[aff1]{Institute for Astronomy and Astrophysics T\"ubingen, Germany}
\affil[aff2]{Max-Planck Institute for Nuclear Physics, Germany}
\affil[aff3]{Universit\'e Paris Diderot, CNRS/IN2P3, France}
\affil[aff4]{University of New South Wales, Australia}
\affil[aff5]{Universit\'e Montpellier, CNRS/IN2P3, France}
\affil[aff6]{University of Amsterdam, Netherlands}

\corresp[cor1]{Corresponding author: gottschall@astro.uni-tuebingen.de}

\maketitle

\begin{abstract}
Supernova remnants (SNRs) are prime candidates for efficient particle acceleration up to the knee in the cosmic ray particle spectrum. In this work we present a new method for a systematic search for new TeV-emitting SNR shells in 2864 hours of H.E.S.S. phase I data used for the H.E.S.S. Galactic Plane Survey. This new method, which correctly identifies the known shell morphologies of the TeV SNRs covered by the survey, HESS\,J1731-347, RX\,1713.7-3946, RCW\,86, and Vela Junior, reveals also the existence of three new SNR candidates. All three candidates were extensively studied regarding their morphological, spectral, and multi-wavelength (MWL) properties. HESS\,J1534-571 was associated with the radio SNR candidate G323.7-1.0, and thus is classified as an SNR. HESS\,J1912+101 and HESS\,J1614-518, on the other hand, do not have radio or X-ray counterparts that would permit to identify them firmly as SNRs, and therefore they remain SNR candidates, discovered first at TeV energies as such. Further MWL follow up observations are needed to confirm that these newly discovered SNR candidates are indeed SNRs.
\end{abstract}

\section{Introduction}
In the past ten years of operation a significant part of observation time of H.E.S.S. has been used to observe the Galactic plane, either survey observations or dedicated pointing observations of selected targets. H.E.S.S. with its large field-of-view of $\sim$\SI{3}{\degree} diameter and a point spread function of $\sim$\SI{0.07}{\degree} is well suited to study extended sources in our galaxy \citep{2006ApJ...636..777A}.
From the currently known Galactic \SI{}{\TeV} $\gamma$-ray sources, the largest fraction consists of still unidentified objects, many of them pulsar wind nebula (PWN) candidates. The work presented here deals with the question whether unidentified or new \SI{}{TeV} $\gamma$-ray in the H.E.S.S. phase I data can be identified as SNRs. Rather than focusing on their spectral properties or possible associations with nearby molecular clouds, our approach is based on a systematic search for shell-like morphologies in extended sources present in the HGPS.
\section{Search method}
The HGPS is a survey of the inner part of the Milky Way. In total, \SI{2864}{\hour} of good quality data are used \citep{Deil:2015}. A sensitivity of at least 2\% of the Crab nebula flux is reached in the covered area and the angular resolution is $\sim$\SI{0.07}{\degree}. On the data products of this survey a systematic search for new TeV-emitting SNR shells is done. 
To search for sources with a shell morphology, a projected 3D shell model emitting homogeneously between $R_{\mathrm{in}}$ and $R_{\mathrm{out}}$ (the shell morphology hypothesis) is tested against a symmetric Gaussian model (the null hypothesis). This is done on a grid of sky coordinates with a spacing of \SI{0.02}{\degree}$\times$\SI{0.02}{\degree}. Due to the large number of test positions, a limited set of parameters, both for the shell and the Gaussian, are tested instead of leaving the parameters free (see Table \ref{tab:parameters}). Therefore, the search is not complete, but we are presenting a systematic approach with a minimal bias coming from the limited set of parameters. The search reproduces all known shell-like sources within the survey region. In addition, three sources showing a shell like morphology are identified.
In order to overcome some of the statistical shortcomings of the gridded search presented above, a source-by-source study on the resulting candidates is performed. For this analysis, additional data that became available beyond the HGPS data set for HESS\,J1534-571 are included. Again, a symmetric Gaussian model and a 3D shell are tested, now letting all parameters to vary freely. Since the models are not nested, the Akaike information criterion \citep{Akaike:1974} was used to assess the significance of the shell over the Gaussian null-hypothesis model.

\begin{table}[h]
\caption{List of tested parameters; shell width $w$ is defined as $w = R_{\mathrm{out}} - R_{\mathrm{in}}$.}
\label{Table:gridparameters}
\centering
\begin{tabular}{lc}
\hline
\multicolumn{2}{c}{Shell ($H_1$) parameters} \\
\hline
$R_{\mathrm{in}}$ & \SI{0.1}{\degree}, \SI{0.2}{\degree}, \SI{0.3}{\degree}, \SI{0.4}{\degree}, \SI{0.5}{\degree}, \SI{0.6}{\degree}, \SI{0.7}{\degree}, \SI{0.8}{\degree} \\
width $w$         & $10^{-5} \times R_{\mathrm{in}}$, $0.1 \times R_{\mathrm{in}}$, $0.2 \times R_{\mathrm{in}}$\\
\hline
\multicolumn{2}{c}{Gaussian ($H_0$) parameters} \\
\hline
$\sigma$          & \SI{0}{\degree}, \SI{0.05}{\degree}, \SI{0.1}{\degree}, \SI{0.2}{\degree}, \SI{0.4}{\degree} \\
\hline
\label{tab:parameters}

\end{tabular}
\end{table}

\section{Dedicated analysis}
\subsection{Sky maps}
From the measured excess and the expected $\gamma$-rays, calculated as
\begin{equation}
N_{ref} = T\int_{E_{\mathrm{min}}}^{\infty}\Phi_{\mathrm{ref}}\left(E\right)A_{\mathrm{eff}}\left(E,q\right)\mathrm{d}E
\end{equation}
assuming a power law differential spectrum ($\Phi(E)$), the surface brightness is derived. In this formula $T$ is the run livetime and $A(E,q)$ the instrument's effective area for a given energy $E$ and observation condition $q$ (e.g.\, zenith angel). 

 The maps are correlated with a circle of radius \SI{0.1}{\degree} and smoothed with a Gaussian filter ($\sigma = $\SI{0.01}{\degree}). The excess is derived using a gamma/hadron separation based on boosted decision trees, a reconstruction technique based on Hillas parameters, and a ring background method\citep{Aharonian:2006, Ohm:2009}. To increase the angular resolution and to reduce the background level, a cut on individual image amplitudes of 160\,p.e. is applied; the resulting energy threshold of the data sets is $\sim$\SI{600}{GeV}.

\subsection{Morphology study}
\renewcommand{\arraystretch}{1.4} 

\begin{table}[h]
\caption{Results from the morphological study of the three new TeV shells. $p_{\mathrm{shell}}$ is the null hypothesis probability that the fit improvement of the shell ($H_1$) over the Gaussian ($H_0$) is due to fluctuations, according to the \textit{Akaike Information Criterion} \citep{Akaike:1974}. Shell fit results: ($l_0$, $b_0$) are the center coordinates, $R_{\mathrm{in}}$ and $R_{\mathrm{out}}$ are the inner and outer radii of the homogeneously emitting spherical shell. }
\label{Table:morphologyresults}

\centering
\begin{tabular}{rrrr}
\hline
& \vtop{\hbox{\strut HESS}\hbox{\strut J1534$-$571}}	& \vtop{\hbox{\strut HESS}\hbox{\strut J1614$-$518}}	& \vtop{\hbox{\strut HESS}\hbox{\strut J1912$+$101}} \\
\hline
$p_{\mathrm{shell}}$  & $5.9 \times 10^{-3}$            & $3.1 \times 10^{-6}$            & $1.7 \times 10^{-6}$           \\
$l_0$$^b$                 & \SI{323.70}{\degree}$^{\SI[retain-explicit-plus]{+0.02}{\degree}}_{\SI{-0.02}{\degree}}$
							 & \SI{331.47}{\degree}$^{\SI[retain-explicit-plus]{+0.01}{\degree}}_{\SI{-0.01}{\degree}}$
							                                   & \SI{44.46}{\degree}$^{\SI[retain-explicit-plus]{+0.02}{\degree}}_{\SI{-0.01}{\degree}}$ \\
$b_0$                 & \SI{-1.02}{\degree}$^{\SI[retain-explicit-plus]{+0.03}{\degree}}_{\SI{-0.02}{\degree}}$
							 & \SI{-0.60}{\degree}$^{\SI[retain-explicit-plus]{+0.01}{\degree}}_{\SI{-0.01}{\degree}}$
							                                   & \SI{-0.13}{\degree}$^{\SI[retain-explicit-plus]{+0.02}{\degree}}_{\SI{-0.02}{\degree}}$ \\
$R_{\mathrm{in}}$     & \SI{0.28}{\degree}$^{\SI[retain-explicit-plus]{+0.06}{\degree}}_{\SI{-0.03}{\degree}}$  
                                                            & \SI{0.18}{\degree}$^{\SI[retain-explicit-plus]{+0.02}{\degree}}_{\SI{-0.02}{\degree}}$  
                                                                                              & \SI{0.32}{\degree}$^{\SI[retain-explicit-plus]{+0.02}{\degree}}_{\SI{-0.03}{\degree}}$ \\
$R_{\mathrm{out}}$    & \SI{0.40}{\degree}$^{\SI[retain-explicit-plus]{+0.04}{\degree}}_{\SI{-0.12}{\degree}} $ 
                                                            & \SI{0.42}{\degree}$^{\SI[retain-explicit-plus]{+0.01}{\degree}}_{\SI{-0.01}{\degree}}$ 
                                                                                              & \SI{0.49}{\degree}$^{\SI[retain-explicit-plus]{+0.04}{\degree}}_{\SI{-0.03}{\degree}}$ \\
\hline
\end{tabular}
\end{table}

\renewcommand{\arraystretch}{1} 
\subsection{Spectra}
In order to obtain an energy spectrum with a broad energy range, the standard procedure is to lower the image amplitude cut. For HESS\,J1534-571 and HESS\,J1614-518, the image amplitude cut is lowered to 60\,p.e., resulting in an energy threshold of $\sim$\SI{300}{\GeV}. In the case of HESS\,J1912+101, this standard procedure, could not be performed however, mostly due to mostly due to problems arising from background estimates which become relevant for large, low surface brightness sources in certain sky areas. Here, the more conservative image cut of 160\,p.e. is kept to reduce systematic uncertainties. The on-source counts for the spectra are extracted from circles with radii $R = R_{\mathrm{out}} + R_{68}$ ($R_{68}$ is the 68\% containment radius of the point spread function, typically \SI{0.07}{\degree}) and centred at the respective best fit positions of the shells. Power laws are fitted to the spectrum of each source ,respectively; more complex models are statistically not favoured.

\begin{table*}[h]
\caption{Spectral fit parameters and results. Both statistical and systematic errors are given for the fit parameter.
}
\label{Table:spectralresults}
\centering
\begin{tabular}{lllll}
\hline
Source & $\mathrm{E}_{\mathrm{min}}$& $\mathrm{E}_{\mathrm{max}}$& $I_{0,1\,\mathrm{TeV}}$ & $\Gamma$   \\
\hline
& \multicolumn{1}{c}{\SI{}{TeV}} &\multicolumn{1}{c}{\SI{}{TeV}} & \multicolumn{1}{c}{\SI{}{\per\centi\square\metre\per\second\per\TeV}} &   \\
\hline
HESS J1534$-$571 & 0.422 & 61.897 & $(2.99\pm0.30\pm0.90)\times 10^{-12}$  & $2.51 \pm 0.09 \pm 0.20$ \\
HESS J1614$-$518 & 0.316 & 38.312 & $(8.33\pm0.49\pm2.50)\times 10^{-12}$ & $2.42 \pm 0.06 \pm 0.20$ \\
HESS J1912$+$101 & 0.681 & 61.897 & $(3.89\pm0.45\pm1.17)\times 10^{-12}$ & $2.56 \pm 0.09 \pm 0.20$ \\
\hline
\end{tabular}
\end{table*}

\begin{figure}[h!]
  {\includegraphics[height=170pt]{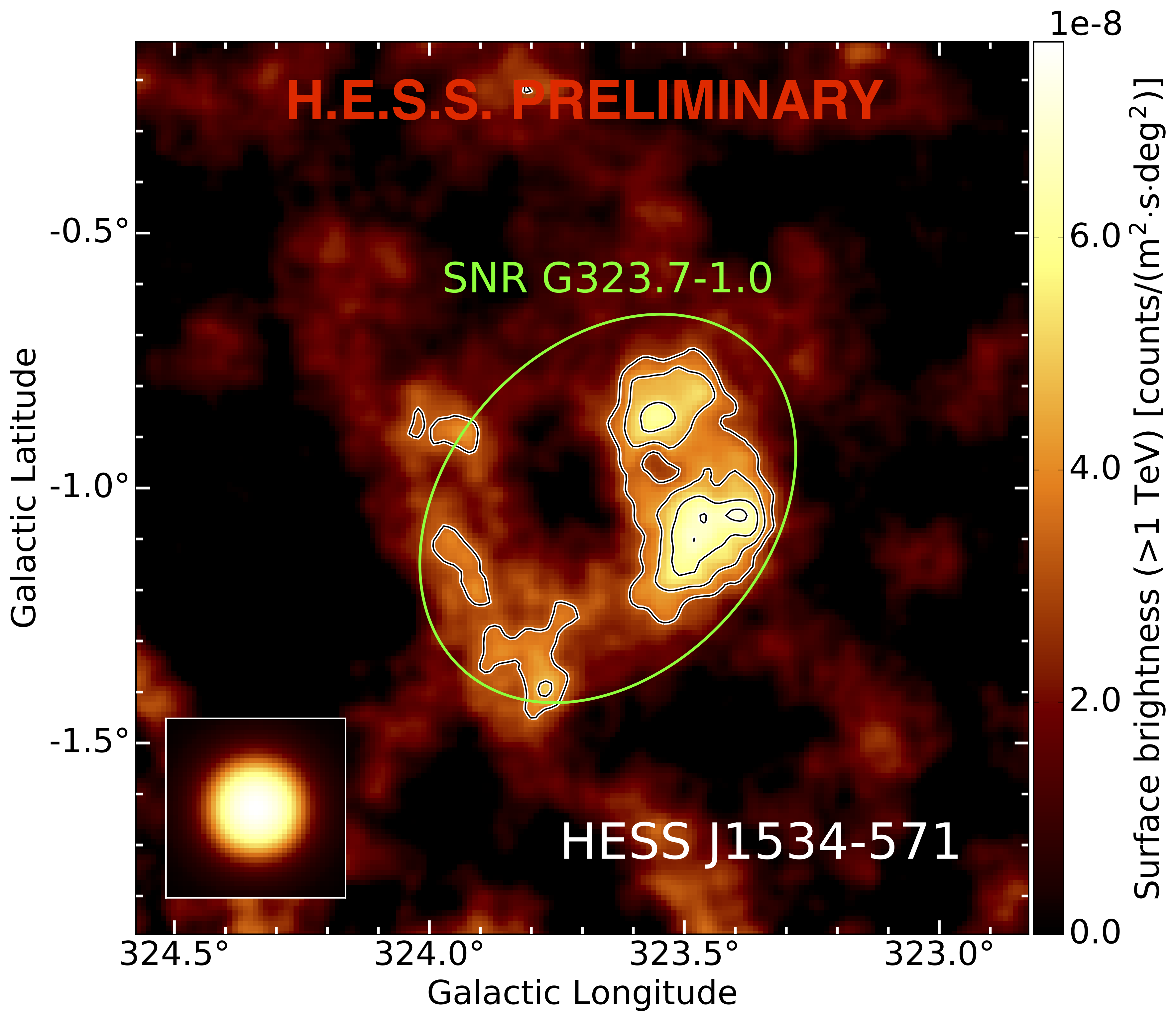}} \hspace{5mm}
  {\includegraphics[height=160pt]{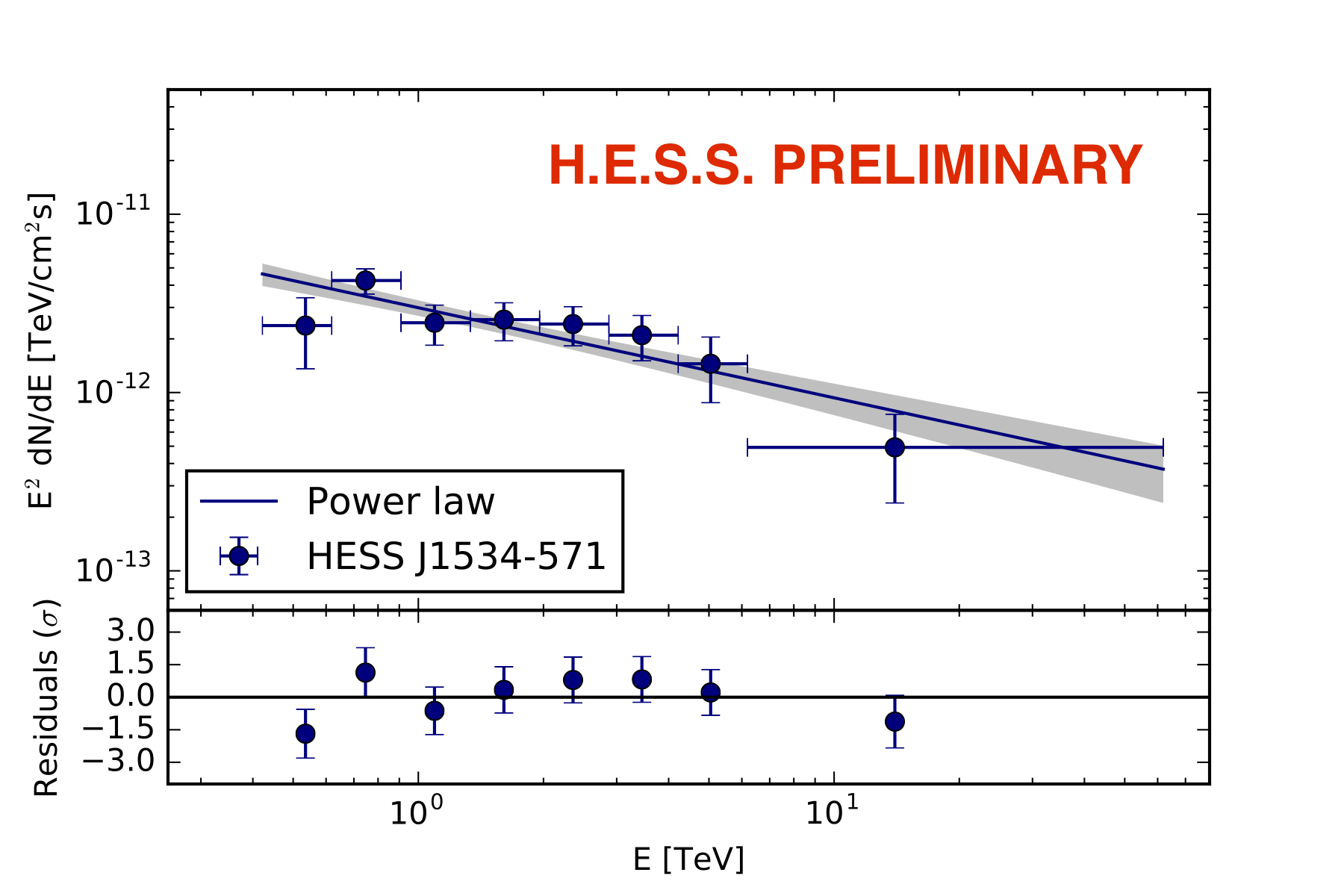}} 
  \caption{Left: Surface brightness map with a correlation radius of \SI{0.1}{\degree} and a Gaussian smoothing ($\sigma=$\SI{0.01}{\degree}; significance contours $3, 4, 5, 6 \sigma$; green ellipse is showing SNR\,G323.7-1.0 \citep{2014PASA...31...42G}. Right: Spectral results with statistical errors fitted with a power law model with $1 \sigma$ error butterfly.}
  \label{fig:plots}
\end{figure}
\begin{figure}[h!]
  {\includegraphics[height=160pt]{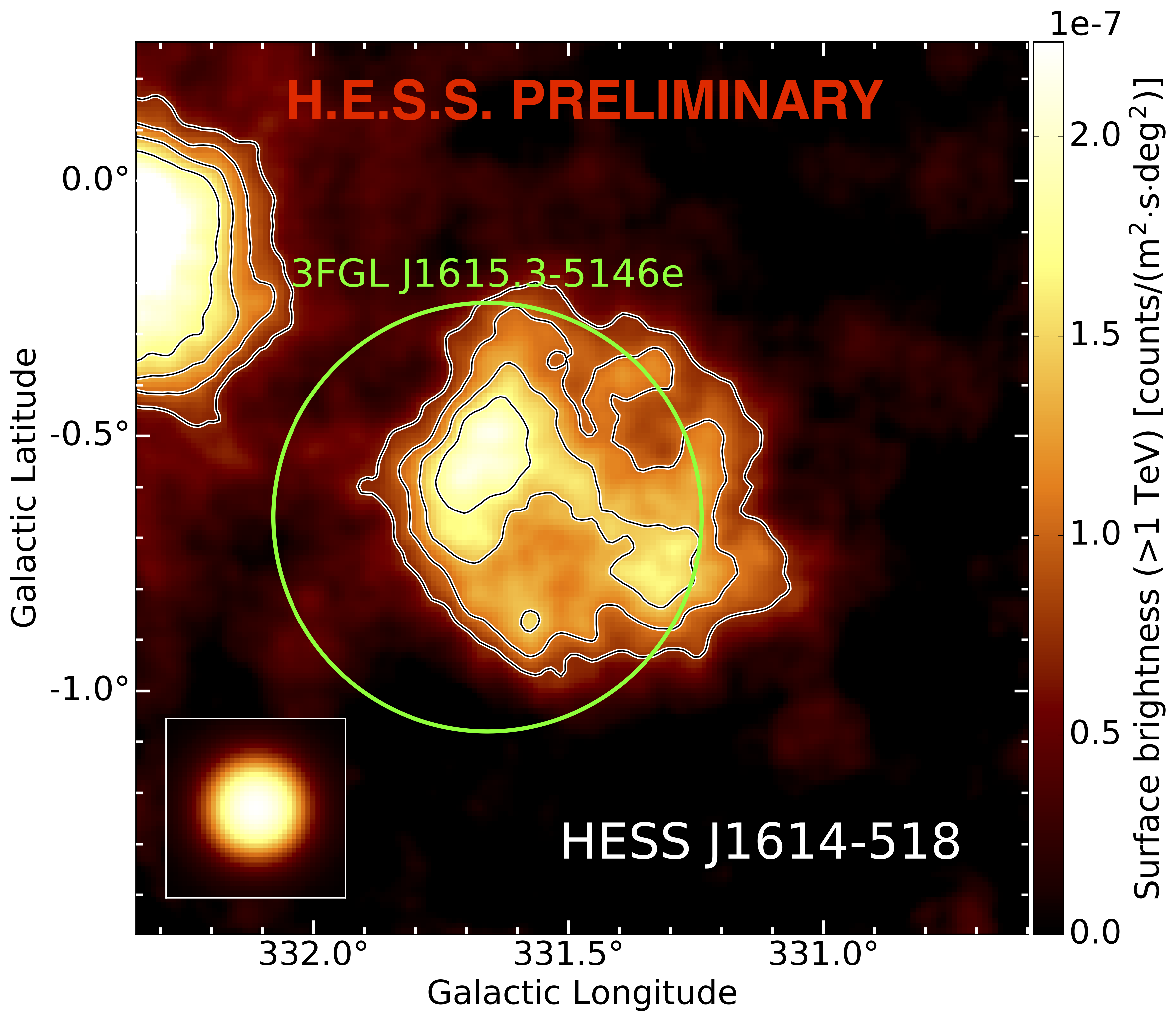}} \hspace{5mm}
  {\includegraphics[height=160pt]{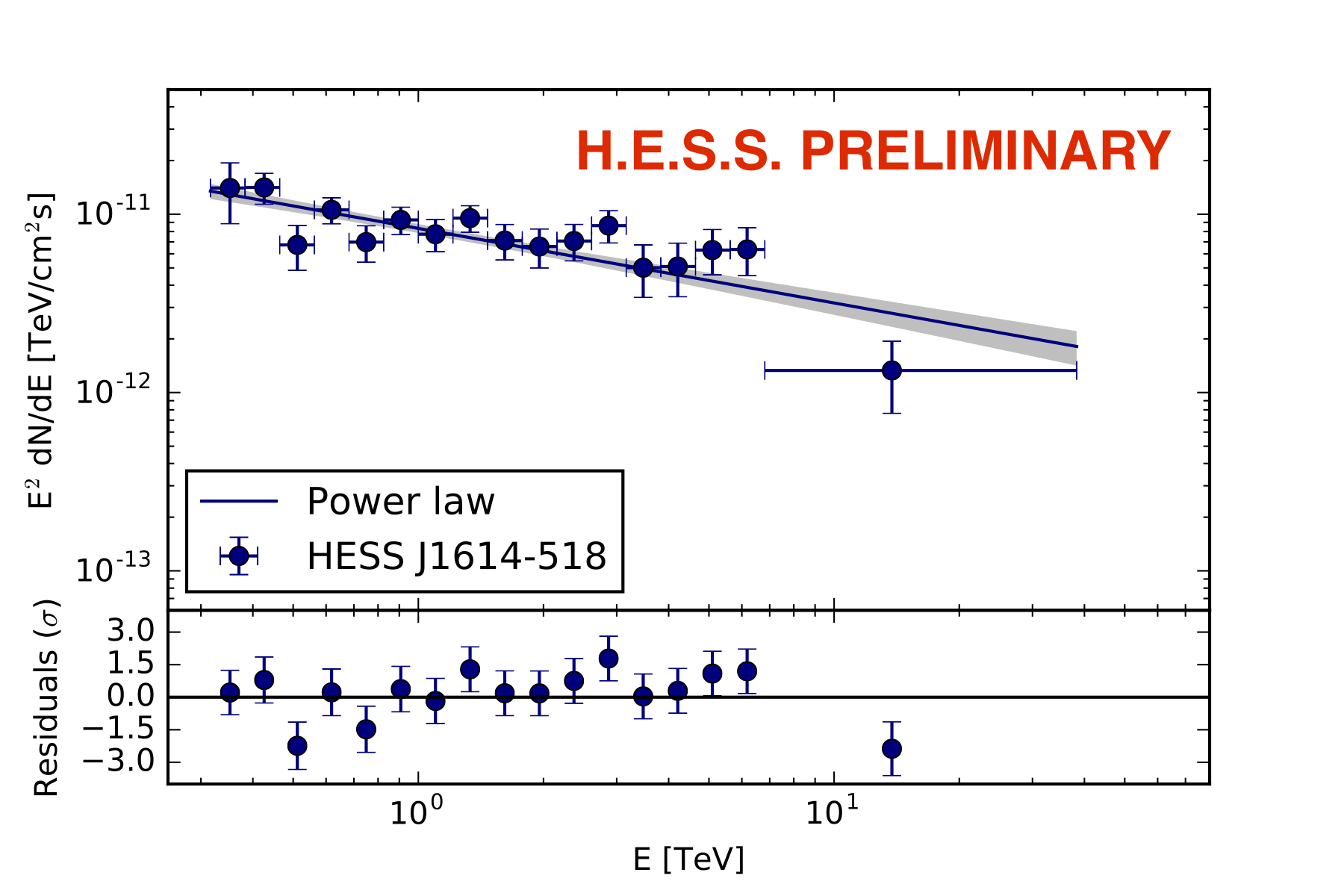}}

  \caption{Left: Surface brightness map with a correlation radius of \SI{0.1}{\degree} and a Gaussian smoothing ($\sigma=$\SI{0.01}{\degree}; significance contours $5, 7, 9, 11 \sigma$; green circle is showing 3FGL\,J1615.3-5146e \citep{2015ApJS..218...23A}. Right: Spectral results with statistical errors fitted with a power law model with $1 \sigma$ error butterfly.}
  \label{fig:J1614}
\end{figure}
\begin{figure}[h!]
  {\includegraphics[height=160pt]{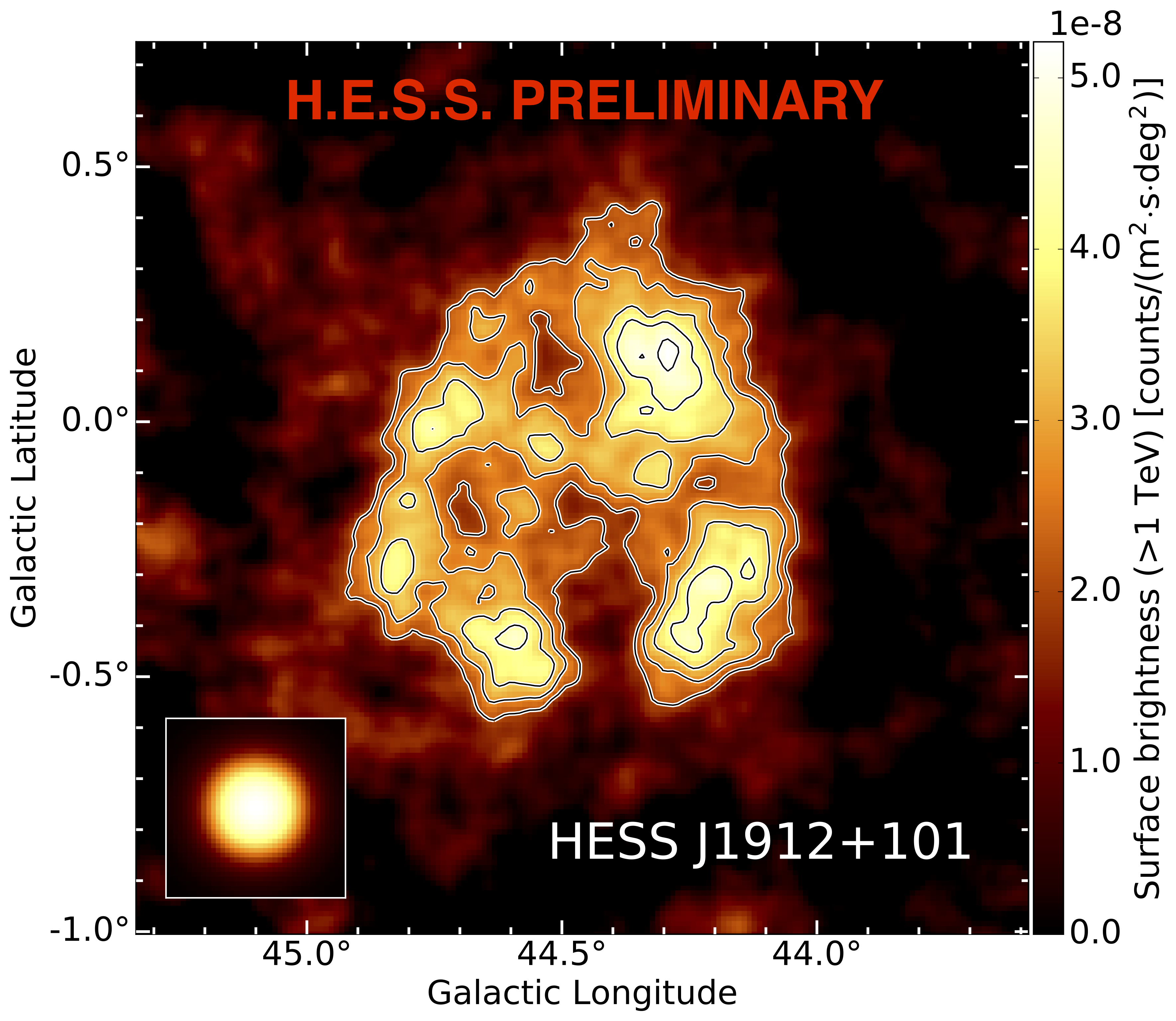}} \hspace{5mm}
  {\includegraphics[height=160pt]{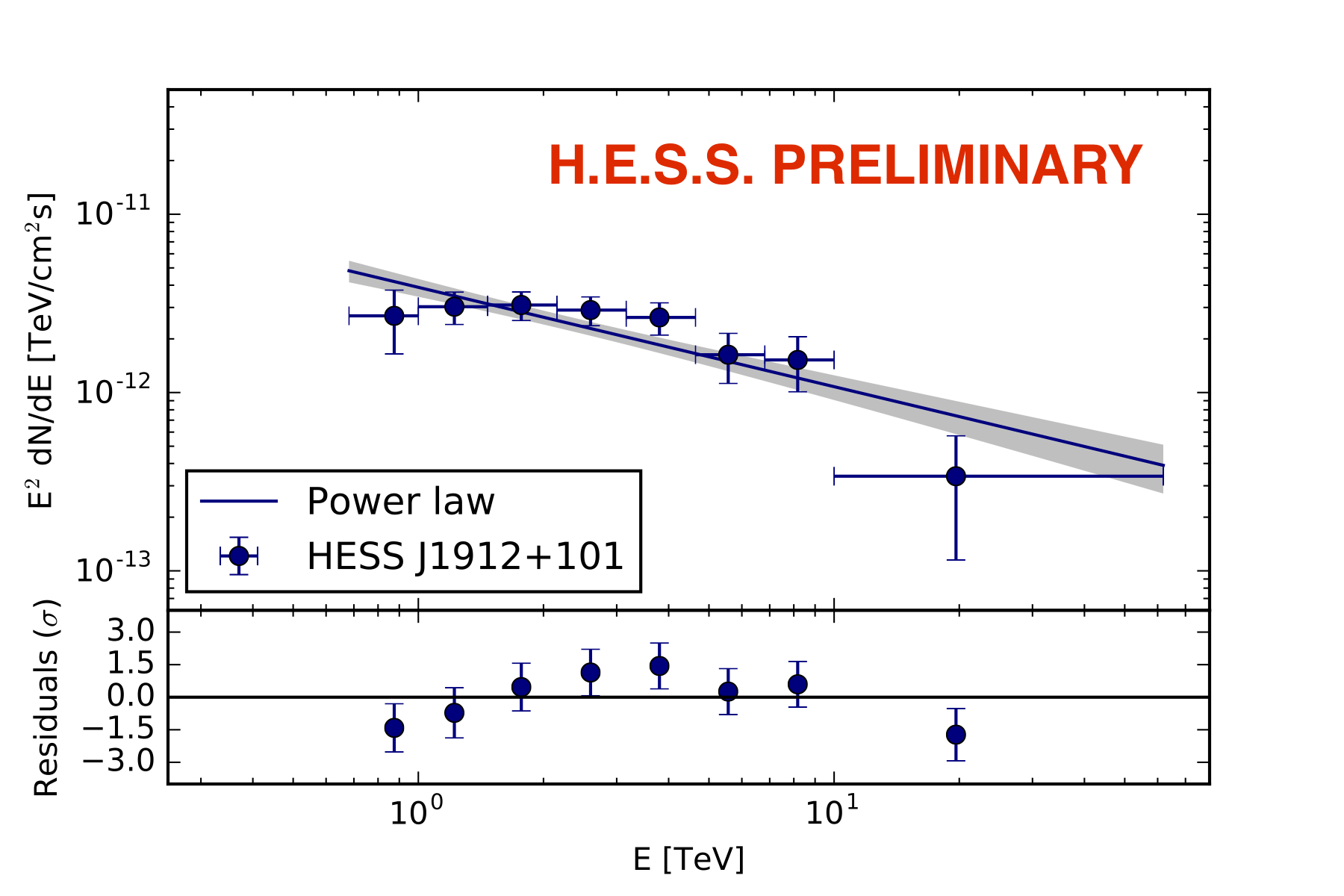}}

  \caption{Left: Surface brightness map with a correlation radius of \SI{0.1}{\degree} and a Gaussian smoothing ($\sigma=$\SI{0.01}{\degree}; significance contours $3,4,5,6,7 \sigma$. Right: Spectral results with statistical errors fitted with a power law model with $1 \sigma$ error butterfly.}
  \label{fig:J1912}
\end{figure}

\section{Mulit-wavelength study}
An SNR candidate of the MGPS2 radio survey is identified as a counterpart of HESS\,J1534-571, from the matching position and shell morphology. Therefore, HESS\,J1534-571 is classified as an SNR. The radio counterpart is represented in the surface brightness map by a green ellipse \citep{2014PASA...31...42G}. 
The Fermi catalogues 3FGL and 2FHL list an extended source at the position of HESS\,J1614-518 (green circle in the surface brightness map) \citep{2015ApJS..218...23A, 2016ApJS..222....5A}. The spectral properties are consistent with the results presented here.
Current observations of HESS\,J1534-571 with Suzaku-XIS have not revealed any X-ray emission from this source. Matsumoto et al. \citep{2008PASJ...60S.163M} report an extended X-ray emission coincident with the North-Eastern component of HESS\,J1614-518.
Because of the lack of SNR or SNR candidate counterparts, HESS\,J1614-518 and HESS\,J1912+101 are classified as SNR candidates.

\section{Conclusions}
The work presented here shows that the current generation of ground-based $\gamma$-ray instruments is well suited to detect new SNRs from a blind search in a survey observation. As a result of this study we classify HESS\,J1534-571 as an SNR and HESS\,J1614-518 and HESS\,J1912+101 as SNR candidates. The absence of a clear non-thermal X-ray emission from these sources might stem from the fact the that $\gamma$-ray radiation observed at very high energies is caused by hadronic interactions. Nevertheless, leptonic scenarios cannot be ruled out at present. Further MWL observations will help to constrain the radiation processes.

\section{ACKNOWLEDGMENTS}
The support of the Namibian authorities and of the University of Namibia in facilitating the construction and operation of H.E.S.S. is gratefully acknowledged, as is the support by the German Ministry for Education and Research (BMBF), the Max Planck Society, the German Research Foundation (DFG), the French Ministry for Research, the CNRS-IN2P3 and the Astroparticle Interdisciplinary Programme of the CNRS, the U.K. Science and Technology Facilities Council (STFC), the IPNP of the Charles University, the Czech Science Foundation, the Polish Ministry of Science and Higher Education, the South African Department of Science and Technology and National Research Foundation, the University of Namibia, the Innsbruck University, the Austrian Science Fund (FWF), and the Austrian Federal Ministry for Science, Research and Economy, and by the University of Adelaide and the Australian Research Council. We appreciate the excellent work of the technical support staff in Berlin, Durham, Hamburg, Heidelberg, Palaiseau, Paris, Saclay, and in Namibia in the construction and operation of the equipment. This work benefited from services provided by the H.E.S.S. Virtual Organisation, supported by the national resource providers of the EGI Federation.

This research has made use of Gammapy, a community-developed, open-source Python package for gamma-ray astronomy \citep{2015arXiv150907408D}.


\nocite{*}
\bibliographystyle{aipnum-cp}%
\bibliography{sample}%

\end{document}